# Reconfigurable Manipulation of Sound with a Multi-material 3D Printed Origami Metasurface

Dinh Hai Le[1,2,*], Felix Kronowetter[2], Yan Kei Chiang[1], Marcus Maeder[2], Steffen Marburg[2], David A. Powell[1,†]

[1] University of New South Wales, School of Engineering and Technology, Northcott Drive, Canberra, ACT 2600, Australia.

[2] Chair of Vibroacoustics of Vehicles and Machines, Department of Engineering Physics and Computation, TUM School of Engineering and Design, Technical University of Munich, Garching b., 85748 München, Germany.

Email: [*]hai.le@unsw.edu.au and [†]david.powell@unsw.edu.au

***Abstract:***

The challenge in reconfigurable manipulation of sound waves using metasurfaces lies in achieving precise control over acoustic behavior while developing efficient and practical tuning methods for structural configurations. However, most studies on reconfigurable acoustic metasurfaces rely on cumbersome and time-consuming control systems. These approaches often struggle with fabrication techniques, as conventional methods face limitations such as restricted material choices, challenges in achieving complex geometries, and difficulties in incorporating flexible components. This paper proposes a novel approach for developing a reconfigurable metasurface inspired by the Kresling origami pattern, designed for programmable manipulation of acoustic waves at an operating frequency of 2000 Hz. The origami unit cell is fabricated using multi-material 3D printing technology, allowing for the simultaneous printing of two materials with different mechanical properties, thus creating a bistable origami-based structure. Through optimization, two equilibrium states achieve a reflection phase difference of $\pi$ through the application of small axial force, ***F,*** or torque, ***T***. Various configurations of the metasurface, generated from different combinations of these two equilibria, enable distinct reflective behaviors with switchable and programmable functionalities. The principle of this work simplifies the shaping of acoustic waves through a straightforward mechanical mechanism, eliminating the need for complex control systems and time-consuming adjustments. This innovative approach paves a novel and effective perspective for developing on-demand switchable and tunable devices across diverse fields, including electromagnetics, mechanics, and elastics, leveraging multi-material printing technology.

1. **Introduction**

Shaping sound waves is central to acoustic engineering as it enables diverse applications ranging from audible to ultrasonic frequencies [1–3]. Mapping the successful control of electromagnetic (EM) waves using metamaterials and their 2D platform, metasurfaces, to the acoustic domain offers a powerful building block in manipulating sound due to the extraordinary features of these artificial materials [4]. The main unique characteristics of acoustic metamaterials and metasurfaces (AMMs) are their lightweight, sub-wavelength thickness, easy fabrication, and their ability to have tailored mass density and bulk modulus at various frequency ranges [4–7]. The use of AMMs to manipulate the propagation of sound waves in unconventional directions raises significant interest due to their potential applications in medical treatment [8–10], acoustic imaging [11], wireless communication [12,13], and sound isolation [14–17].

Recent advances in manipulating sound using AMMs lie in developing real-time reconfigurable or function-programmable structures and devices [18]. Compared to employing piezoelectric [19,20] and magnetic materials [21,22] for creating electromagnetically active designs, altering the physical configuration of AMMs proves to be a more practical and powerful technique. Nevertheless, conventional AMMs have typically been produced using Fused Deposition Modeling (FDM) 3D printing technology [23,24], which poses challenges in creating reconfigurable structures. Common approaches to address these challenges include using actuators connected to solid parts of the structures [13,25–27], manually arranging numerous selective fixed elements with varying reflection phase shifts [28,29], or controlling the propagating medium within the static structures [30–32]. Another promising solution to enhance the reconfigurability of AMMs using the FDM technique is the utilization of flexible filaments or shape memory materials [33–35]. Despite these enhancements, using FDM techniques to fabricate AMMs still suffers from disadvantages, including low resolution, limited material options, and difficulty fabricating complicated geometries. With the rapid expansion of additive manufacturing over the past decades, advanced 3D printing techniques like multi-material jetting and stereolithography have emerged as promising fabrication methods for creating reconfigurable acoustic metasurfaces by utilizing flexible and self-changing materials [7,36].

Origami, known for its foldable, deformable, deployable, and collapsible features, offers significant advantages regarding physical configuration and structural adaptability [37]. Beyond its artistic origins, origami research has diversified into various applications of construction [38], energy harvesting [39], EM absorbers [40,41], and robotics [42,43]. In terms of controlling sound waves, origami-based and origami-inspired structures have been used in absorption [44], sound barriers [45], tunable attenuation [46], and acoustic waveguides [47]. The first and only research on origami-inspired reconfigurable acoustic structures for manipulating sound direction indicates that the folded state is controllable, providing various functions of acoustic focusing, splitting, and one-way transmission [48]. However, the proposed origami structures suffer from multiple undesired degrees of freedom, fabricated from flat, rigid materials, which can complicate their construction process and compromise their robustness.

Leveraging advances in additive manufacturing to create origami structures is a promising approach to achieving highly foldable, robust, homogeneous, and lightweight forms. Applying this concept to manipulating acoustic waves presents a novel opportunity to enhance sound-shaping capabilities. In this paper, we numerically and experimentally investigated a bistable origami-based programmable acoustic metasurface. Multi-material 3D printing technology is applied to fabricate the origami unit cell by simultaneously printing two materials with different mechanical features. Two equilibria are switched by left-handed or right-handed twisting the structure, providing different reflection phases, as shown in Figures 1a and 1b. We applied a Newtonian iteration process to optimize the structure's geometry such that the phase difference between the two equilibria equals $\pi$, encoding two configurations as bits "0" and "1". By mechanically programming an array of origami units using different coding sequences, the reconfigurable metasurface achieves beam focusing and splitting functions at 2000 Hz, as illustrated in Figure 1c. It is worth noting that the measurement setup was meticulously calibrated around 1800–2600 Hz, ensuring the most accurate results at 2000 Hz, which was chosen as the operating frequency to evaluate the design.

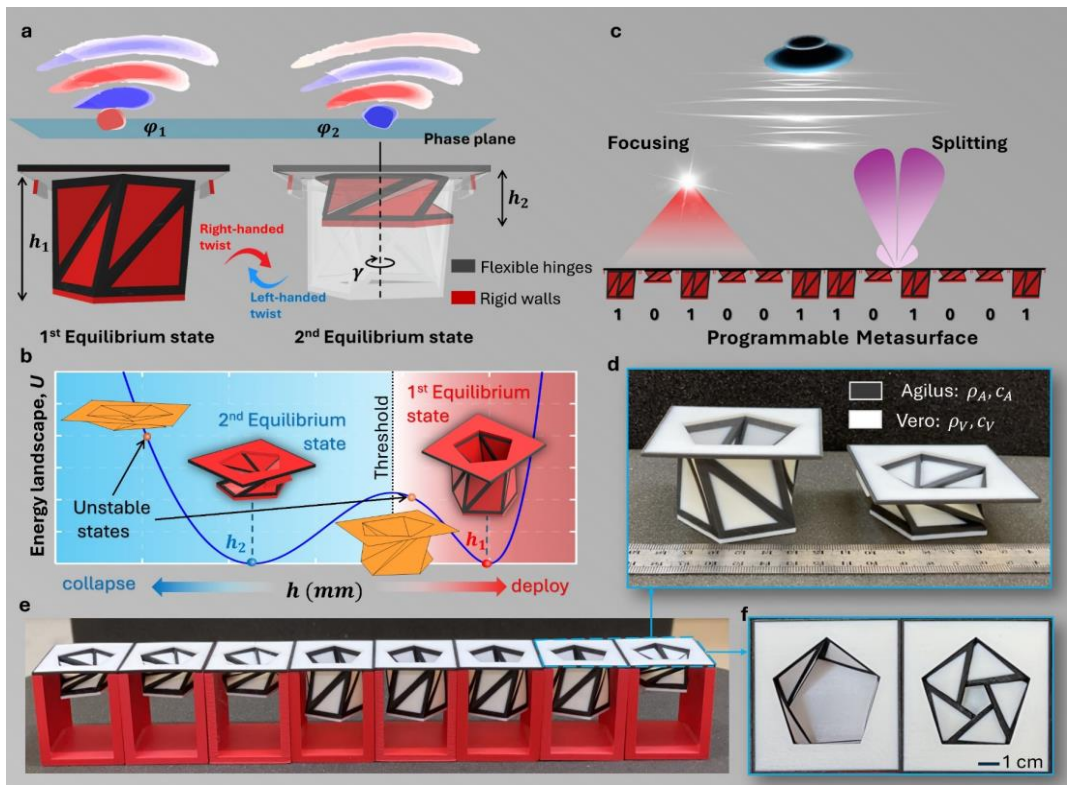

**Figure 1. Conceptual Illustration of the bistable origami-based acoustic metasurface**. a) Operation of a bistable unit cell composed of flexible hinges (black) and rigid walls (red). The two equilibria can be transformed back and forth by a right- or left-handed twist. The height of the first equilibrium state is $h_1$, and that of the second is $h_2$. Each stable state responds to reflecting sound with different phase. b) Illustration of the strain energy landscape variation during the collapse and deployment processes. c) The concept of a reconfigurable metasurface with two programmable functions, beam focusing and beam splitting, by controlling bistable unit cells at either first or second equilibrium states. The first and second equilibrium states are encoded as bits "1" and "0", respectively. d) The 3D-printed origami unit cell at two equilibria with fabricated materials. e) The 3D printed origami-based metasurface sample at the coding sequence 00011110. f) The top view of the first (left) and second (right) equilibrium states.

## 2. Results.

### 2.1. Design and geometry

We employed the Kresling pattern, a well-known non-rigid origami, in the designed structure due to its extensive applications across various fields of science and technology, including miniature robots [42], energy absorption [49], and vibration attenuation [50]. Composed of flexible hinges, which represent the valley and mountain folds, and rigid counterparts, the origami-based structure is collapsible and deployable, leading to the axial displacement (relative height, $h$, and the radius, $r$), and the rotation (twist angle, $\gamma$) about the z-axis. Supplementary Note 1 and the Methods section comprehensively analyze the detailed geometry of the Kresling origami structure and its associated strain energy. We analyzed the folding state by considering the strain energy as the structure deforms, which renders most configurations unstable. During structural transformation, the energy landscape changes nonlinearly, reaching minima at one or two equilibrium points, which enables a monostable or bistable configuration of the structure. It is important to note that this investigation considers only structures where the strain energy goes to zero at all equilibria.

Figure 1d shows the perspective view of 3D-printed origami units at different stable states. We fabricated the unit cells using PolyJet technology, which prints rigid (white) and flexible (dark grey) materials simultaneously. The rigid walls are made from Vero, a rigid plastic material, while the flexible components use Agilus, a rubber-like material [51]. The material of flexible hinges is crucial in enabling deformation, as it replicates the folding behavior of thin paper. It is worth noting that the acoustic impedance of Agilus and Vero is significantly higher than that of air due to their high mass density. Consequently, the surfaces are considered sound-hard in both simulations and experiments. The Methods section provides details on the 3D printing process and material properties. Figure 1e illustrates the fabricated bistable origami-based metasurface in a random configuration, showcasing different combinations of first and second equilibrium states. The metasurface consists of eight identical bistable origami units supported by red 3D-printed frames.

We control the configurations of the metasurfaces by applying small torsional or axial loads to each origami unit. Figure 1f exhibits the top view of two origami units in the first (left) and second (right) equilibrium states. Two stable states can be transformed back and forth by left- and right-handed twists (See Movie S1 in Supplementary Information).

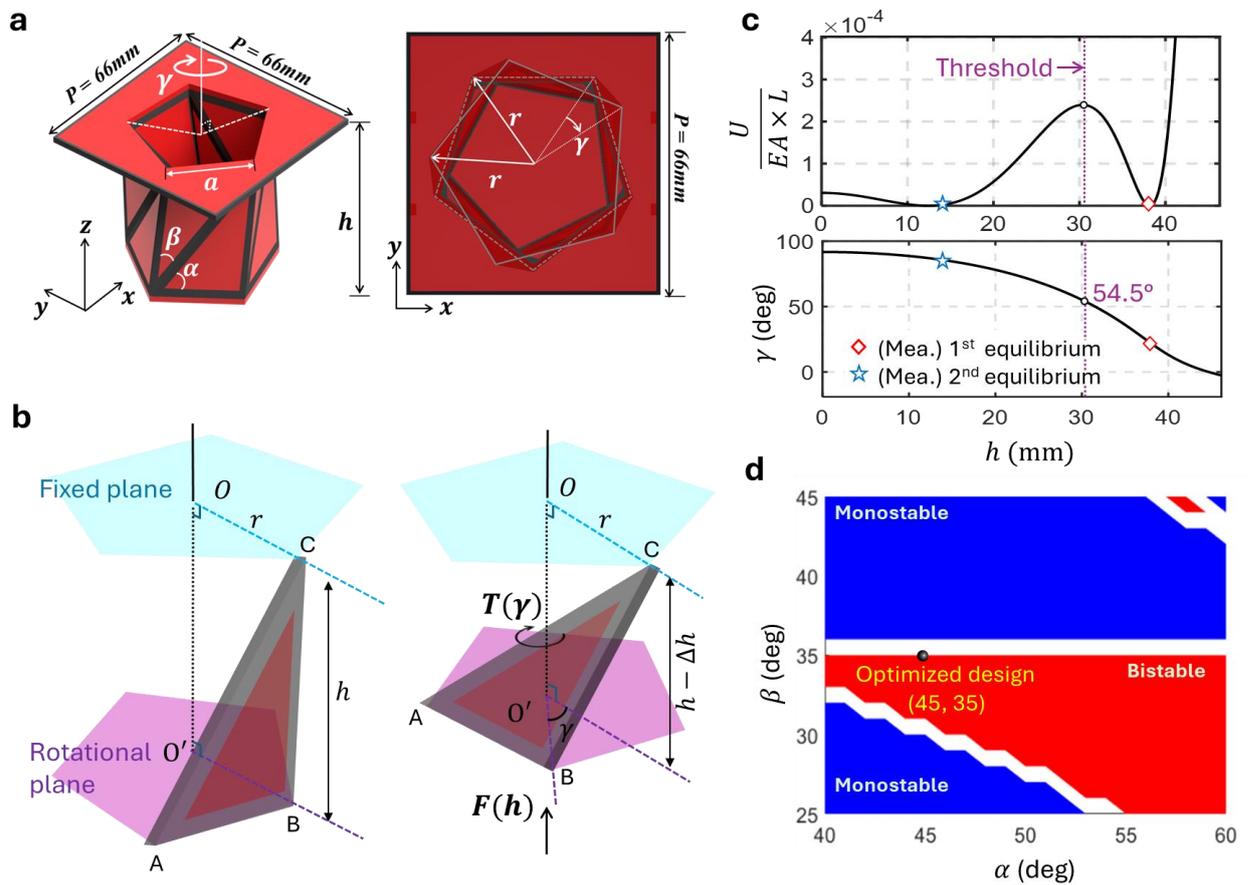

**Figure 2. Structural geometry and mechanical behavior of the origami unit.** a) The perspective and top view of the origami unit with its structural geometry and dimensions. The deformation of the origami unit results from the arrangement of ten identical triangles, each defined by bottom side length $a$ and two angles $\alpha$ and $\beta$. b) Transformation of the coordinates of a single triangle during deformation. c) The energy landscape (top) and twist angle (bottom) are functions of the height, $h$, of the unit cell. The markers indicate that the measured relative height and twist angle result in two equilibrium states. d) Regions of monostable and bistable states vary with changes in $\alpha$ and $\beta$.

Figure 2a illustrates the perspective and top views of the bistable origami unit. The overall dimensions of the unit cell are 66 mm × 66 mm × 42 mm. The structure comprises three layers. The top square layer has a dimension of 66 mm × 66 mm and a thickness of $t_1 = 2$ mm. The bottom layer with a thickness of $t_1 = 2$ mm is attached to the lower end of the middle layer (origami layer). The origami layer consists of 10 identical triangles with a thickness of $t_2 = 1.2$ mm (see details of the three layers in Figure S2a, Supplementary Note 1). Each triangle has a bottom side length of $a = 32$ mm and two angles, $\alpha = 45.1°$ and $\beta = 35°$. The deformation of the origami unit occurs in the middle layer, dominated by the Kresling origami pattern. During the deformation, $\alpha$ and $\beta$ are constant, whereas three variables, $h$, $\gamma$, and radius, $r$, are coupled, defining the folding state (see details in the Experimental Section). The structure remains stable at the first or second equilibrium without an external load. After applying a small force, $F(h)$, or torque, $T(\gamma)$, each triangle element of the origami unit transforms as shown in Figure 2b. During the transformation, the vertical axis $OO'$ and the top layer plane remain unchanged. In contrast, the bottom layer experiences a rotational movement by an angle of $\gamma$ and a displacement within a $\Delta h$ range relative to the $OO'$ axis. The coordinates of three points, $A$, $B$, and $C$, can be determined from the parameters $h$, $\gamma$, and $r$. Figure 2c presents the normalized strain energy and the relationship between the twist angle $\gamma$ and displacement $h$. The normalized strain energy reaches the two minima at $h_1 = 38.2$ mm and $h_2 = 12.4$ mm, suggesting that the theoretical results indicate zero strain energy at the two equilibria. The relation between twist angle and relative height reveals that any change in the twist angle results in a corresponding change in the height of the unit cell. The twist angles at the first and second equilibrium states are 20.3° and 86.5°, respectively. It is worth noting that the threshold twist angle, which marks the transition point between the two equilibrium states, is determined to be $\gamma = 54.5°$. It demonstrates that the origami unit will either collapse or deploy to reach another stable state by twisting the structure's bottom layer beyond this threshold angle. The blue and red markers are the unit cell's measured height and twist angles at two equilibria, showing good agreement with the analytical results. Figure 2d depicts the regions of monostable and bistable states as $\alpha$ and $\beta$ vary within the ranges of $[40°, 60°]$ and $[25°, 45°]$, respectively. The monostable state is in blue, representing the region where the strain energy reaches a minimum (zero) at a single point.

Conversely, the red-colored region indicates the bistable state, where strain energy reaches minima at two points. This figure shows how the monostable and bistable states can be controlled by adjusting the values of $\alpha$ and $\beta$. However, not all combinations of $\alpha$ and $\beta$ result in a $\pi$ phase difference between the two stable states. An optimization process is employed to determine the optimal values of $\alpha = 45.1°$ and $\beta = 35°$, ensuring that the structure achieves the desired phase difference when reflecting the impinging sound (see Supplementary Note 2 and Experimental Section for optimization procedure).

## 2.2. Simulation results

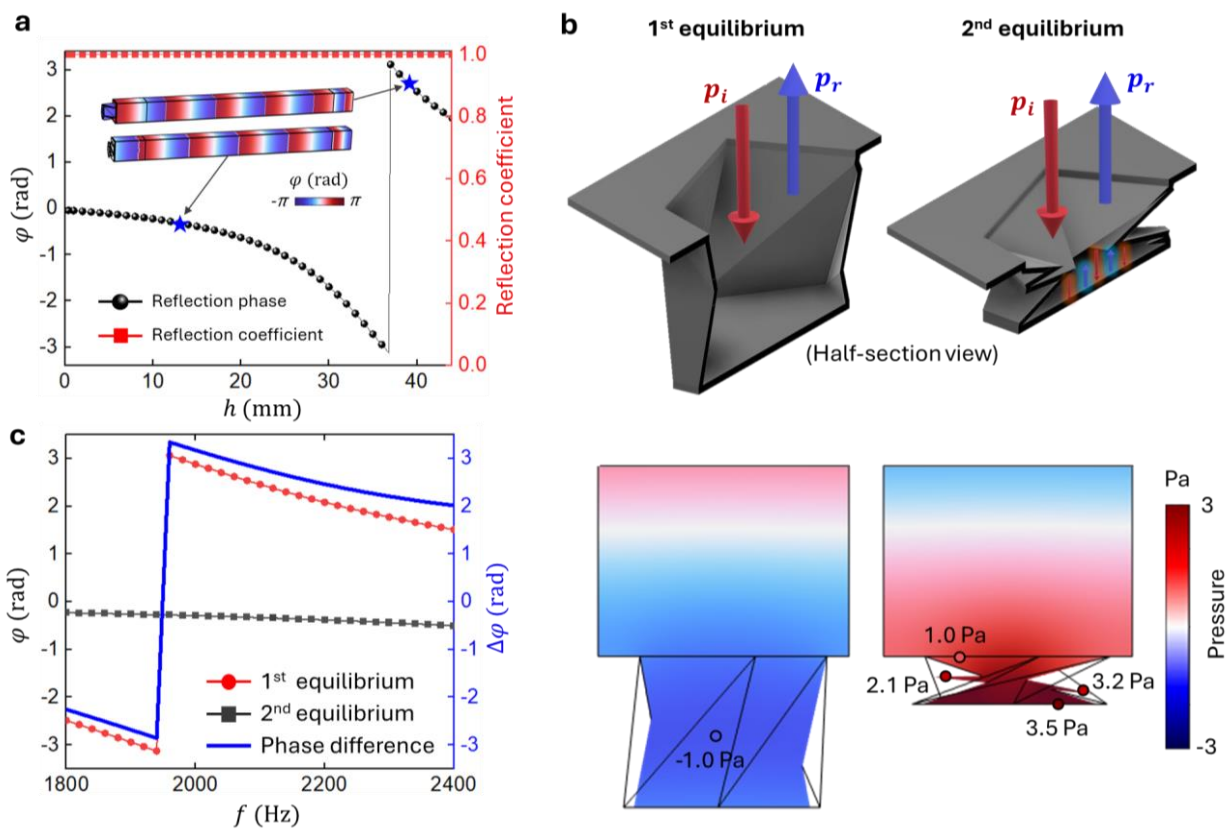

**Figure 3. Simulated acoustical characteristics.** a) Dependence of reflection phase and amplitude on a relative height at 2000 Hz. The inset shows reflection phases at two equilibria. b) (Top) Explanation of acoustic behavior, and (bottom) the acoustic pressure fields at two equilibrium states. c) The reflection phases of the structure at two equilibria and the phase difference vary with changes in operating frequency.

Figure 3a exhibits the origami-based unit cell's simulated reflection phase and amplitude during displacement, $h$. To simplify the modeling and analysis, we considered the origami unit to have negligible thickness and treated it as sound-hard boundaries. The incident field is primarily reflected during deformation, maintaining a 100% reflection coefficient at the first and second equilibrium points. The reflection phase shifts within approximately $2\pi/3$ range as the displacement varies from 0 mm to 45 mm. The reflection phase does not encompass a $2\pi$ phase shift because the maximum depth of the origami unit is a quarter of the wavelength, resulting in the longest propagation path being $\lambda/2$. At the two equilibrium points, the reflection phases differ by $\pi$. The inset shows the 3D result of the reflection phase at both equilibria. Figure 3b presents the scattered pressure fields at 2000 Hz, which helps to understand the physical mechanisms underlying the acoustic behavior at the two equilibria. The amplitude of the acoustic pressure field at the first equilibrium matches that of the incident field, set at 1 Pa. In contrast, the pressure field at the second equilibrium is confined within the structure and exhibits an amplitude of 3.5 Pa, i.e., 3.5 times stronger than that at the first equilibrium. This relation demonstrates that the structure behaves like an acoustic waveguide in the first equilibrium state and functions as a Helmholtz-like resonator in the second equilibrium state. This behavior results from the deformation of the structure in the second equilibrium state. When the structure collapses, it forms a neck and a cavity that traps and amplifies the acoustic pressure field, causing the origami unit function as a Helmholtz resonator. Note that the origami unit confines the pressure field to its bottom half and the gaps between the folded triangles. Figure 3c illustrates the reflection phase of two stable states working with different frequencies from 1800 Hz to 2400 Hz. The $\pi$ phase difference occurs only at 2000 Hz. The reflection phase of the first equilibrium state is more sensitive to frequency variations than the second equilibrium state, which maintains greater consistency. It is worth noting that the effect of thermos-viscous losses is insignificant in this investigation, which we verify through numerical comparison in Supplementary Note 3.

## 2.3. Experimental verification

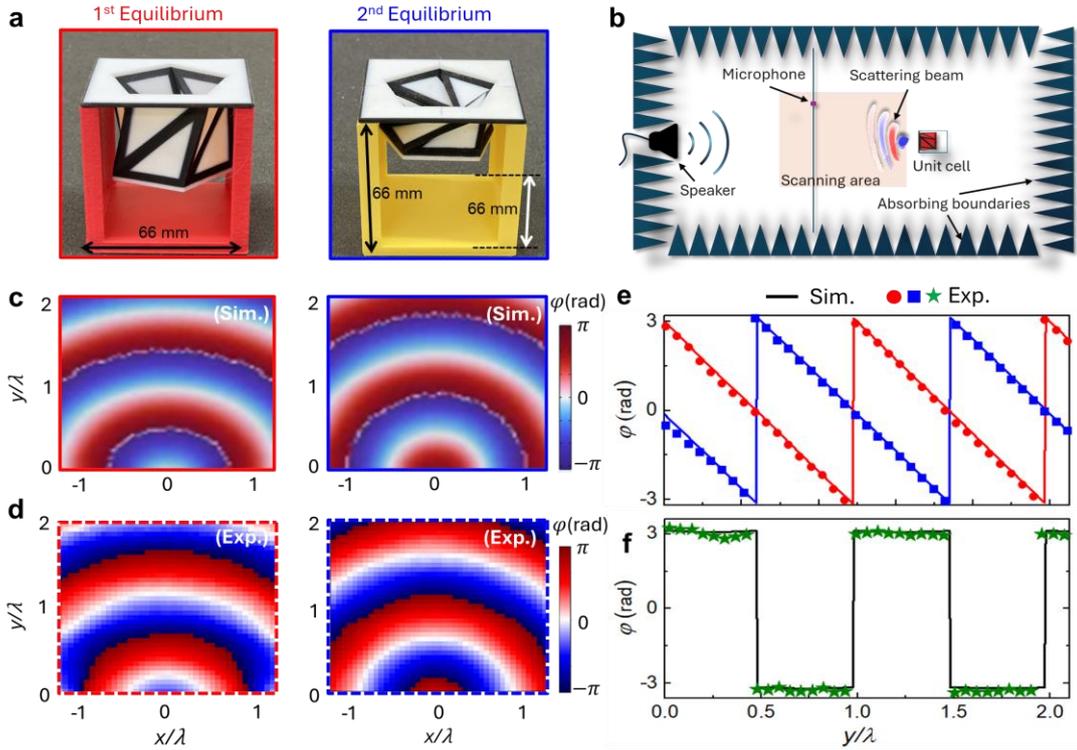

**Figure 4**. **Experimental verification of single unit cell**. a) 3D printed sample of the proposed unit cell at first (left) and second (right) equilibrium states. Red and yellow printed supports have dimensions of 66 mm × 66 mm × 66 mm. b) Schematic of measurement set up for a single unit cell in a parallel-plate acoustic waveguide. Comparison between c) simulation and d) measurement of the 2D scattered field from a single unit cell at 2000 Hz. e) Comparison between simulated and measured data along vertical cutline at position $x = 0$. f) The $\pi$ phase difference between two equilibria.

Before verifying the performance of the programmable metasurfaces, we experimentally investigated the acoustic behavior of a single-unit cell in both equilibrium states within a parallel plate waveguide. We designed a 3D-printed frame with dimensions of 66 mm × 66 mm × 66 mm to support the origami unit, ensuring its stability within the parallel plate waveguide that has a height of 66 mm, as depicted in Figure 4a. Furthermore, Figure 4b provides the schematic diagram of the experimental setup in the parallel plate waveguide apparatus. A single speaker is placed at the center front of the waveguide, radiating a normal sound wave at 2000 Hz. We position a microphone between the unit cell and the speaker, allowing it to move continuously along the x and y axes to record the reflected signal within a 400 mm × 360 mm area. We placed the unit cell 800 mm away from the speaker and fill the apparatus boundary with absorbing foam to eliminate unwanted reflections (See [Experimental Section](Experimental Section) for a detailed measurement process). Figures 4c and 4d compare the simulated and measured reflection phases at the two equilibrium states, respectively. The simulation setup replicates the same dimensions as the experiment, and the captured data from the same measurement area show good agreement with the experimental results in both stable states. Figures 4e and 4f present the extracted simulation and measurement data along the y-axis at the 400 mm center vertical cutline between the speaker and the unit cell. The results depict an excellent match between the simulation and the measurement data. Notably, the phase difference between the two stable states is $\pi$ in the simulation and measurement.

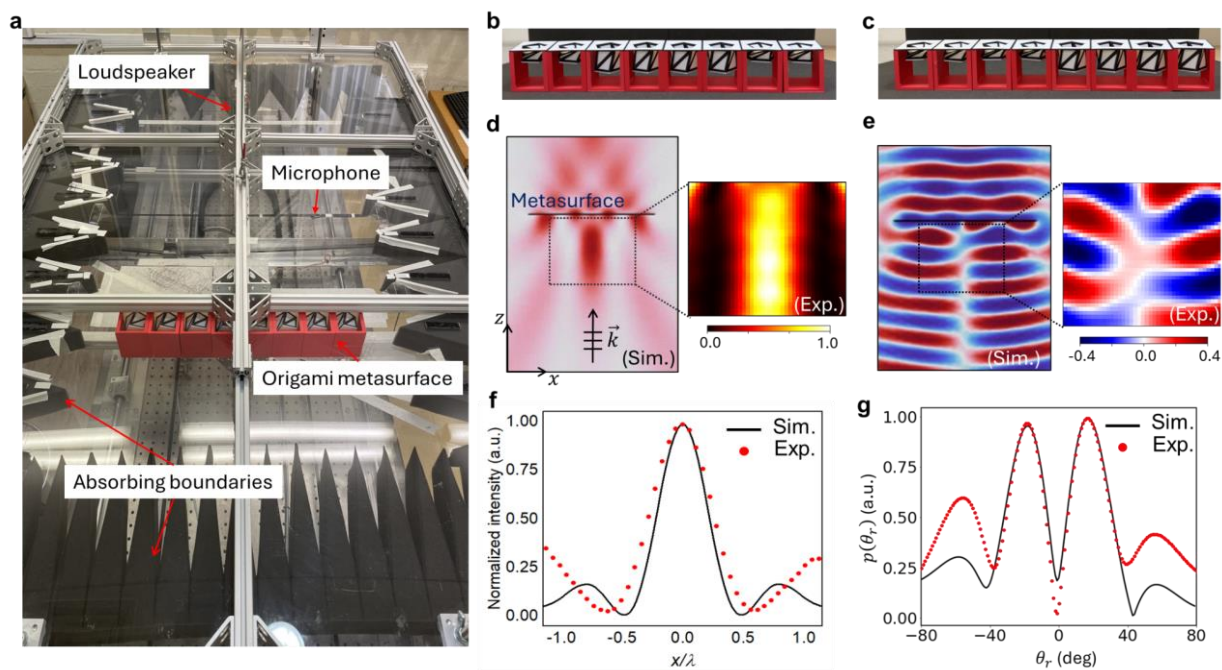

**Figure 5**. **Experimental verification of the programmable metasurface**. a) Photograph of the experimental setup using a parallel plate waveguide. The configuration of programmable metasurface at coding sequences of b) 00111100 (for focusing function), and c) 00001111 (for splitting function) Comparison between simulation and measurement of the metasurface at functionalities of d) beam focusing and e) beam splitting. Comparison of f) the normalized sound intensity and g) far-field scattering pattern between simulation and measurement.

We numerically and experimentally investigate a programmable metasurface with multifunctional capabilities. The metasurface is a combination of different unit cells configured at either the first or second equilibrium state, which provides $\varphi_1$ rad and $\varphi_2 = \varphi_1 + \pi$ rad phase shift. For easy visualization, we encode the first equilibrium as bit "1" and the second as bit "0". Arranging the origami units in different coding sequences enables distinct functionalities, suggesting a readily programmable metasurface. The combination of unit cells with reflected phase difference of $\pi$ rad to manipulate wave propagation is well reported in the electromagnetic domain [52–54] and has been applied to control sound waves [55,56]. This technique is powerful for creating digital coding metasurfaces to actively control sound for various functions. However, achieving $\pi$ rad phase shift in a single acoustic structure is challenging as it typically lacks reconfigurability, requires complicated control circuits [57], or demands a complicated optimization process [57,58].

In this paper, beam focusing and beam splitting are the targeted functionalities. As the apparatus size is limited, the metasurface consists of eight origami units, having a total length of 528 mm, as illustrated in Figure 5b and Figure 5c. A 3D-printed frame supports each origami unit to help it stand stable within the waveguide. Figure 5a is a photograph of the experimental setup in the parallel plate waveguide, similar to Figure 4b, but with the origami unit cell replaced by a metasurface. The origami units of the metasurface are manually reconfigured between two equilibrium states to achieve the target function. The coding sequence of the metasurface for the focusing function is 00111100, with the reflection phase distribution of each origami unit adhering to the Generalized Snell's Law (GSL), expressed as:

$$\varphi = \varphi_0 + \frac{2\pi}{\lambda}\left(\sqrt{(x-x_0)^2 + (y-y_0)^2}\right), \qquad (1)$$

where $(x_0, y_0)$ represents the focal point position with the initial reflection phase $\varphi_0$. The coding sequence for the splitting function is 11110000 (or 00001111), derived from the far-field scattering pattern. The normalized far-field scattered pressure of the finite metasurface, $p(\theta_r)$, results from:

$$p(\theta_r) = \int_{-\infty}^{\infty} p_s e^{ik_0 \sin \theta_r l} dl, \qquad (2)$$

where $p_s$ is the complex scattered acoustic pressure extracted along a horizontal line with a specified length, $l$ and i is the imaginary unit. Furthermore, $k_0$ denotes the wave number and $\theta_r$ represents the elevation angle of the acoustic beam direction. Figures 5b and 5c display photographs of the fabricated metasurface, illustrating these two functions. The reconfigurability of each function is discussed in Supplementary Notes 4 and 5. Figure 5d compares the focusing behavior of the metasurface between simulation and measurement. These results agree that the metasurface can act as a sound-focusing lens with a focal length of 65 mm. Figure 5f compares the normalized sound intensity along the x-axis at the focal point in the simulation and measurement, showing a substantial similarity. Figures 5e and 5g depict the metasurface's second target function, beam splitting. Good agreement between simulation and measurement of scattered pressure fields is shown in Figure 5e, while Figure 5f compares the far-field scattered pressure. With the coding sequence 11110000, the metasurface provides two split acoustic beams at a steering angle of 20º, demonstrating a solid agreement between simulation and experiment.

## 3. Conclusion

This work numerically and experimentally investigates an origami-based reconfigurable acoustic metasurface capable of achieving multiple functions, such as tunable focusing lenses and steerable beam splitting. The metasurface comprises flexible origami units, 3D printed using multi-material printing technology. This technique allows for the simultaneous printing of rigid and flexible materials, leading to robust and readily reconfigurable structures. The origami-based unit can be collapsed or deployed by applying left-handed or right-handed twisting or an axial load, achieving two equilibrium states. Following an optimization process, the proposed origami structure demonstrates optimized stable configurations that reflect sound waves with a $\pi$ phase shift difference at an operating frequency of 2000 Hz. The first and second stable states are binary encoded as bits "1" and "0", corresponding to the reflection phase of $\varphi$ rads and $(\varphi + \pi)$ rads. We programmed the metasurface, consisting of eight binary-switchable origami units, into multiple configurations, each with various coding sequences in combinations of "0" and "1". Each metasurface exhibited distinct reflective acoustic behavior. The principle of this work makes controlling acoustic waves easier based on a simple switch mechanism without the need for complex control systems or time-consuming adjustments. This approach brings numerous possibilities for applying origami building blocks and advanced 3D printing technology to develop reconfigurable acoustic and electromagnetic metasurfaces.

## 4. Experimental section

### *4.1. Numerical simulation*

We numerically simulated the origami unit and the metasurface using the finite-element method-based 3D full-wave simulation from the commercial software package COMSOL Multiphysics 6.2 with the pressure acoustic module. The background pressure field is an air medium with a mass density of $1.12 \text{ kg m}^{-3}$ and sound speed, of $343 \text{ m s}^{-1}$. The measured reflection phase of the proposed origami unit aligns with that of the rigid 3D-printed counterpart, which is characterized by its acoustically rigidity with mass densities of $1180 \text{ kg m}^{-3}$ and sound speeds of $2260 \text{ m s}^{-1}$ (see details in [Supplementary Note 7](#)). Consequently, we modelled the deformable origami unit with zero thickness, treating it as sound-hard boundaries in the numerical simulation. Perfectly matched layers are placed at the boundaries of the simulation domain to avoid unwanted reflections. In the simulation, the setup aligns with the experimental dimensions of the parallel plate waveguide. We utilized COMSOL-MATLAB Livelink to support the optimization process, where the reflection phase from COMSOL and the relative height and twist angle from computations in MATLAB are interconnected. This integration allows for the computation of the phase differences between two equilibria across many parametric simulations.

## *4.2. Optimization*

### 4.2.1. Deformation analysis

To understand the mechanical behavior during the deformation process, we calculated the strain energy stored in the structure through an equivalent rigid truss model [59,60]:

$$U(h,r,\gamma) = \frac{5EA}{2}(L_{AB}.\varepsilon_{AB}^2 + L_{BC}.\varepsilon_{BC}^2 + L_{AC}.\varepsilon_{AC}^2), \tag{3}$$

where $EA$ is the axial rigidity of the truss elements, $L_{AB}$, $L_{BC}$, $L_{AC}$ are the length of line $AB$, $BC$, $AC$, and $\varepsilon_{AB}$, $\varepsilon_{BC}$, $\varepsilon_{AC}$ are their corresponding strains (see details in Supplementary Note 1). As these strains can be controlled by axial force, $F(h)$, and torque, $T(\gamma)$, allowing the origami unit to deform through rotation by angle $\gamma$, vertical translation $\Delta h$, or horizontal displacement $\Delta r$. This paper selects the height, $h$, as prescribed, while $r$, and $\gamma$ are dependent variables. By considering different conditions of axial changes with free-rotation or torsional loading with free horizontal displacement, the equilibrium states, as well as the relation between $r$, $h$, $\gamma$, and $U$, can be determined by solving the following nonlinear Equation:

$$\left.\frac{\partial U(h,r,\gamma)}{\partial \gamma}\right|_h = \left.\frac{\partial U(h,r,\gamma)}{\partial r}\right|_h = 0, \quad \left.\frac{\partial^2 U(h,r,\gamma)}{\partial \gamma^2}\right|_h > 0, \quad \left.\frac{\partial^2 U(h,r,\gamma)}{\partial r^2}\right|_h > 0, \tag{4}$$

where $\partial U(h,r,\gamma)/\partial r|h$ and $\partial U(h,r,\gamma)/\partial \gamma|h$ denote the restoring force and torque during deformation.

### 4.2.2. Newton iteration method

We applied a Newtonian Iteration method to optimize the structural parameters, starting with the initial values of $\alpha = 50°$ and $\beta = 32.5°$, selected arbitrarily in the bistable region presented in Figure 2d. Equation (4) determines the values of $h$, $\gamma$, and $r$ of two equilibria for each set of parameters of $\alpha$ and $\beta$. These values were input into the COMSOL-MATLAB Livelink to calculate the reflection phase difference between two stable states ($\Delta\varphi_{\text{sim}}$). We compared the calculated phase difference with the target phase difference of $\Delta\varphi_{\text{target}} = \pi$. The optimization process was convergent when phase error, expressed as $||\Delta\varphi_{\text{sim}}| - \Delta\varphi_{\text{target}}|$, was less than the predefined error tolerance of 0.01 rad. The optimized values of $\alpha$ and $\beta$ are 45º and 35º, respectively. Figure S3, Supplementary Note 2 illustrates the optimization process and phase error throughout the iterations.

### 4.3. Fabrication

We initially designed eight flexible origami units in Autodesk Inventor Professional 2023 with the geometry shown in Figure S2 and Table S1. We then fabricated these origami units using the PolyJet (material jetting) 3D printing technology with a multi-material printer Stratasys J750 (Stratasys Ltd.), which features a resolution of 14 $\mu$ m layer height and 40 $\mu$ m in the x- and y-axes. The bimaterial used consists of Agilus, a rubber-like material with a Shore A hardness of 30 for flexible hinges, and VeroMagentaV, with a Shore D hardness of 85, for rigid walls. A gel-like material (SUP706B) was employed during printing and subsequently removed using a Waterjet cleaning system. The structures dried naturally at room temperature. The 3D support frames were fabricated from polylactic acid (PLA) using Fused Deposition Modeling (FDM) technology with a Prusa 3.9 3D printer.

### 4.4. Measurement

The experimental verification was conducted in a parallel plate waveguide, surrounded by absorbing materials, with a height of 66 mm, allowing only a single propagation mode at operating frequencies below 2600 Hz. We used a loudspeaker to generate the normally incident sound field and carried out measurements of single-unit cells and metasurfaces with the sample placed 80 cm away from the sound source. The complex values of total acoustic pressure were recorded by a microphone attached to a moveable belt, using point-by-point measurements (10 mm step) within an area of $40 \times 36 \text{ cm}^2$ ($2.3\lambda \times 2\lambda$). We calculated the scattered pressure field and reflection phase by subtracting the incident field from the total field.

**Acknowledgments**

The authors acknowledge the financial support from by Australian Research Council Discovery Project DP200101708, and Universities Australia/DAAD joint research co-operation Scheme Project 57446203.

**Author contributions**

D.H.L and D.A.P conceived the idea, D.H.L conceived the origami metasurface, performed numerical simulation, theoretical analysed, validated optimization, fabricated samples, performed measurement, and wrote original draft. F.K performed simulation and validated optimization analysis. Y.K.C provided guidance on numerical simulation, measurement process and supported in writing. M.M, S.M provided guidance on theoretical analysis and optimization procedures. D.A.P provides guidance on the entire of the work. All authors participated in discussion, review, and approve the manuscript.

**Competing Interests**

The authors declare no competing financial interests.

**Supplementary Information**

Supplementary Information accompanies this manuscript as part of the submission files.

# Supplementary Information for

# Reconfigurable Manipulation of Sound with Multi-material 3D Printed Origami Metasurface

Dinh Hai Le[1,2], Felix Kronowetter[2], Yan Kei Chiang[1], Marcus Maeder[2], Steffen Marburg[2], David. A. Powell[1]

[1] University of New South Wales, School of Engineering and Technology, Northcott Drive, Canberra, ACT 2600, Australia.

[2] Chair of Vibroacoustics of Vehicles and Machines, Department of Engineering Physics and Computation, TUM School of Engineering and Design, Technical University of Munich, Garching b., 85748 München, Germany.

Email: *hai.le@unsw.edu.au and †david.powell@unsw.edu.au

This file includes:

Figure S1-S8

Movie S1: Deformation of a fabricated origami unit

Supplementary Note 1. Kresling origami pattern and 3D-designed geometry

Supplementary Note 2. Optimization procedure

Supplementary Note 3. Influence of thermoviscous acoustics on reflective performance

Supplementary Note 4. Tunable acoustic metalens

Supplementary Note 5. Steerable acoustic splitter

Supplementary Note 6. Comparison of reflection phase between rigid and flexible structures

**Movie S1: Deformation of a fabricated origami unit**

[Supplementary Movie S1.mp4](Supplementary Movie S1.mp4)

**Supplementary Note 1. Kresling origami pattern and 3D designed geometry**

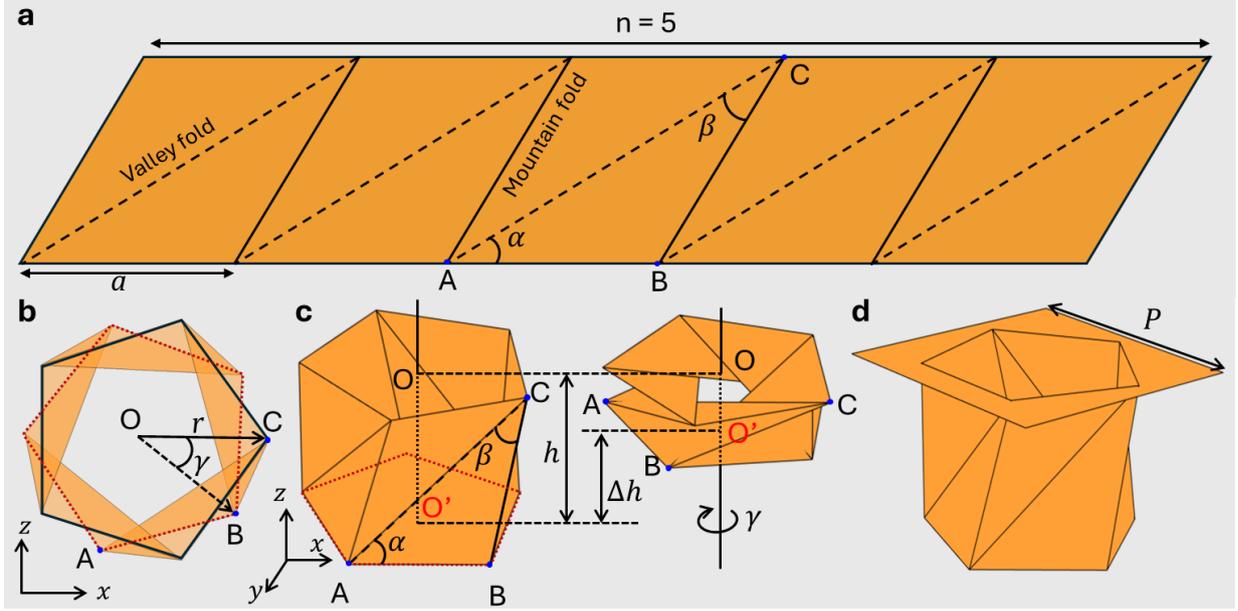

**Figure S1.** Kresling origami pattern and geometry. a) The flat sheet with a crease pattern. c) top view of the folded Kresling origami. c) Deformation of the origami with displacement in z axis. d) The geometry of the proposed origami-based unit cell, neglecting the thickness of the walls.

The proposed structure is Kresling origami pattern, which includes ten identical triangles separated by alternative valley (dashed) and mountain (solid) crease lines, as illustrated in Figure S1. Each triangle is defined by a length $a$, and two angles, $\alpha$ and $\beta$. We consider the transformation of a triangle $ABC$ instead of all ten triangles, as their deformation is identical. In the flat sheet form, the dimensions of the $ABC$ are:

$$L_{AB} = a, \qquad L_{BC} = a\frac{\sin\alpha}{\sin\beta}, \qquad L_{AC} = a\frac{\sin(\alpha+\beta)}{\sin\beta}. \qquad (S1)$$

The folded origami is a pentagonal cylinder ($n = 5$). Figures S1b and S1c indicate the folded origami at the top and perspective views. Three variables determine the folded state: twist angle, $\gamma$, radius, $r$, and height, $h$. During deformation, the coordinates of the axis $OO'$ and triangle $ABC$ are:

$O\,(0,0,0), \qquad O'(0,0,-h)$

$$A\left(r\cos\left(\frac{2\pi}{n}+\gamma\right), -r\sin\left(\frac{2\pi}{n}+\gamma\right), -h\right), \qquad B\left(r\cos\frac{2\pi}{n}, -r\sin\frac{2\pi}{n}, -h\right), \qquad C(0,0,r) \qquad (S2)$$

The folded lengths of the triangle are:

$$l_{AB} = 2r\sin\frac{\pi}{n}, \qquad l_{BC} = \sqrt{h^2 - 2r^2\cos\gamma + 2r^2}, \qquad l_{AC} = \sqrt{h^2 - 2r^2\cos\left(\frac{2\pi}{n}+\gamma\right) + 2r^2}. \qquad (S3)$$

During transformation, the structure will undergo strain. The strains in these folding lines result from:

$$\varepsilon_{AB}(h,r,\gamma) = \frac{l_{AB} - L_{AB}}{L_{AB}}, \qquad \varepsilon_{ABC}(h,r,\gamma) = \frac{l_{BC} - L_{BC}}{L_{BC}}, \qquad \varepsilon_{AC}(h,r,\gamma) = \frac{l_{AC} - L_{AC}}{L_{AC}} \qquad (S4)$$

The strain energy stored in the structure during deformation is:

$$U(h,r,\gamma) = n\frac{EA}{2}(L_{AB}\cdot\varepsilon_{AB}^2 + L_{BC}\cdot\varepsilon_{BC}^2 + L_{AC}\cdot\varepsilon_{AC}^2), \qquad (S5)$$

where $EA$ is the tensile rigidity of the truss. We define the normalized strain energy as:

$$U_{\text{norm}} = \frac{U}{EA \times L}, \qquad (S6)$$

where $L = L_{AB} + L_{BC} + L_{AC}$ is the total length of the triangle $ABC$

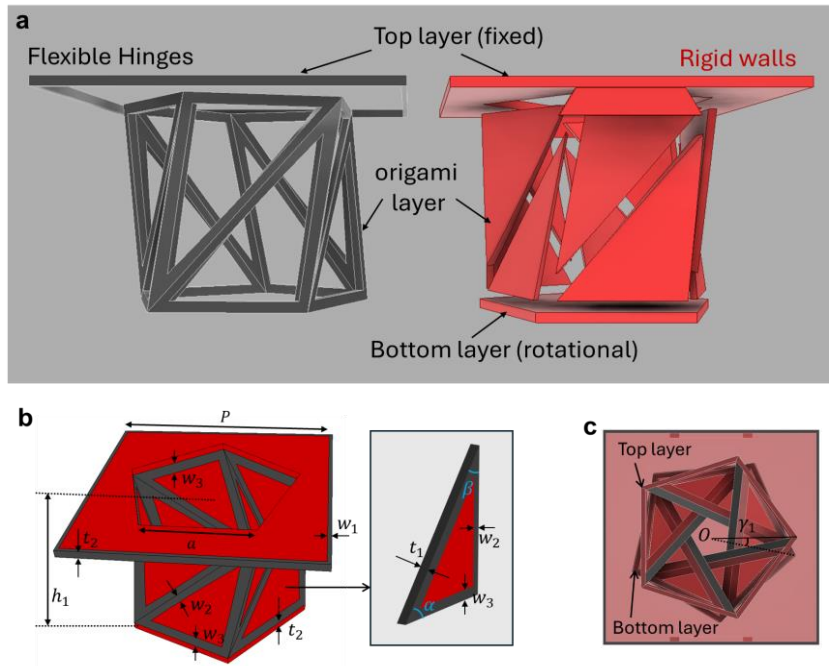

**Figure S2.** 3D-designed origami unit at first equilibrium state and its detailed dimensions. a) Separated parts of origami unit. b) Perspective and c) top views of the origami unit.

Figure S2 shows the 3D-designed origami unit in the first stable state with detailed structural geometries. The structure consists of three layers: a squared top layer, a middle origami layer, and a bottom layer.

Unlike the non-thickness design used in simulations, the full-size 3D structure used for fabrication has specific thicknesses, as described in Figure S2b, with their values listed in Table S1. We designed and fabricated the origami units at first equilibrium.

**Table S1**: Designed parameters of an origami unit cell at first equilibrium state.

| Parameters | $P$ | $a$ | $t_1$ | $t_2$ | $w_1$ | $w_2$ | $w_3$ | $h_1$ | $\gamma_1$ (deg) |
|---|---|---|---|---|---|---|---|---|---|
| Values (mm) | 66 | 32 | 1.2 | 2.0 | 1 | 1.5 | 2.0 | 42.2 | 20.3° |

**Supplementary Note 2. Optimization procedure.**

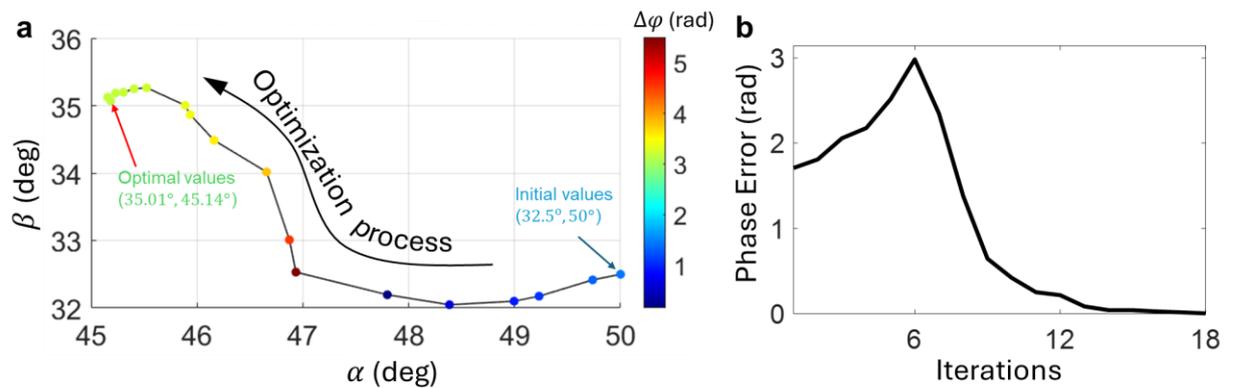

**Figure S3.** Optimization procedure. a) Optimization process for achieving optimal designed values of $\alpha$ and $\beta$. b) Phase error throughout the iterations.

We can adjust the equilibrium points of the Kresling origami by varying the angles $\alpha$ and $\beta$. Each specific pair of these angles generates a distinct phase shift for each stable state, $\varphi_1$, and $\varphi_2$, thereby producing a corresponding phase difference between those two equilibria, $\Delta\varphi = |\varphi_1 - \varphi_2|$. To realize $\Delta\varphi = \pi$, identifying the optimal values of $\alpha$ and $\beta$ is essential. Figure S3 outlines the optimization process employed to determine these optimal values. The process is initiated with $\alpha = 50°$ and $\beta = 32.5°$, converging after 18 iterations to the optimal values of $\alpha = 45.14°$ and $\beta = 35.01°$. For the final design, we rounded these values to $\alpha = 45°$ and $\beta = 35°$. Figure S3b depicts the phase error observed throughout the iterations.

**Supplementary Note 3. Influence of thermos-viscous acoustics on reflective performance**

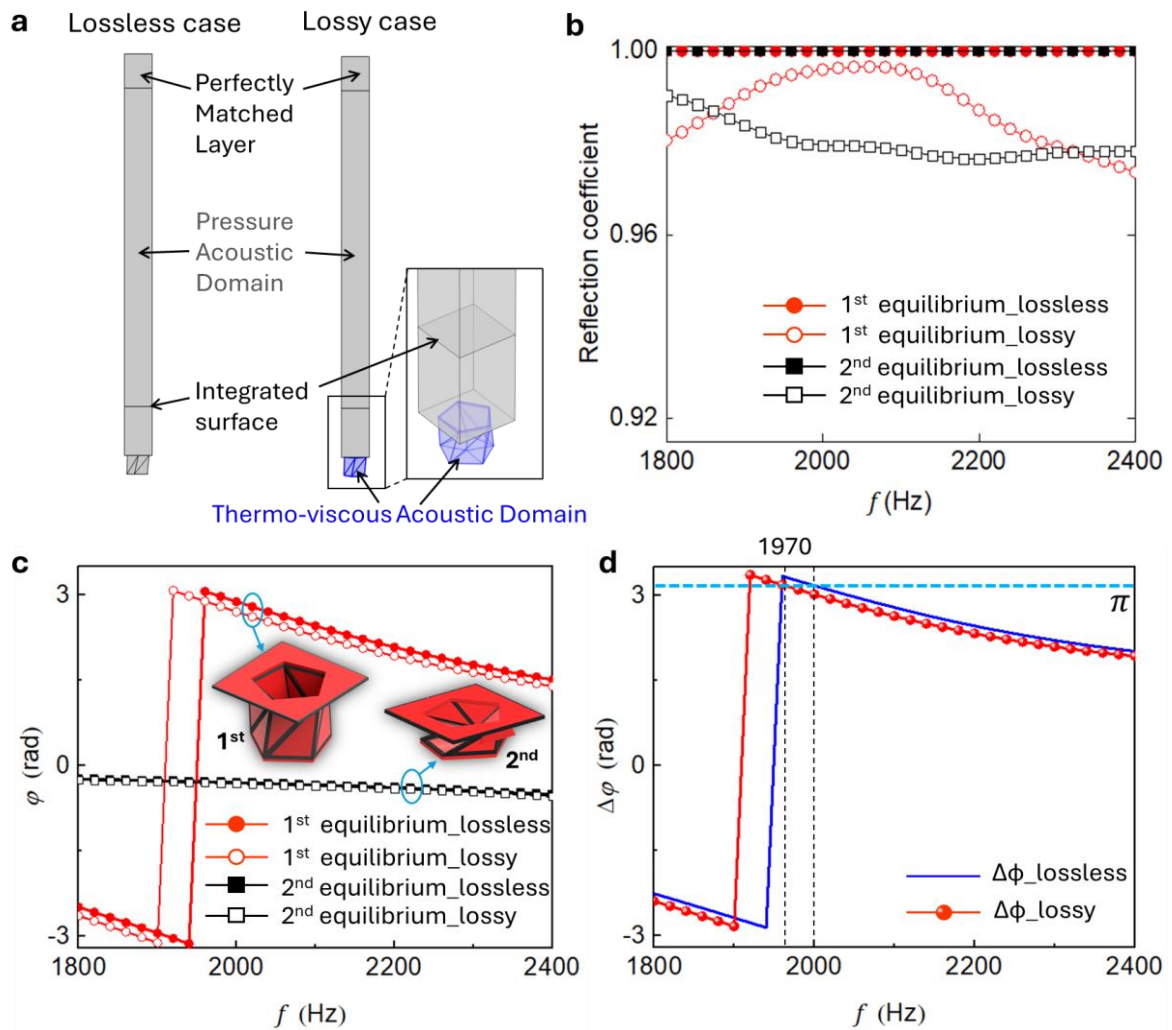

**Figure S4.** Influence of thermos-viscous losses on the reflective performance. a) Simulation setup for a single origami unit comparing the lossless (left) and lossy (right) cases. Comparison of b) reflection amplitude, c) reflection phase, and d) phase difference between two equilibrium states in lossless and lossy conditions.

To investigate the influence of thermos-viscous losses on the reflection phase, we employed the Thermo-viscous Acoustics module in parallel with the Pressure Acoustics module in COMSOL Multiphysics, as shown in Figure S4a. The effect of thermal conduction and viscous friction results in acoustic energy dissipation and attenuation inside the origami unit. Figure S4b shows the impact of thermos-viscous losses on the reflection coefficient compared to the ideal, lossless scenario. The results reveal a slight decrease in reflection under lossy conditions, indicating the minimal influence of the thermos-viscous loss. Notably, the reflection coefficient remains above 97.5 % across all investigated frequencies for both equilibrium states, regardless of whether the conditions are lossless or lossy. Figure S4b illustrates the impact of thermos-viscous losses compared to an ideal lossless medium on the reflection phases of the two equilibria. The thermos-viscous losses significantly affect the reflection phase in the first equilibrium state, while their impact on the second equilibrium state is minimal. This results in a shift of the $\pi$-phase difference between the two equilibria from 2000 Hz under lossless conditions to 1970 Hz in the presence of thermos-viscous losses, as presented in Figure S4c.

**Supplementary Note 4. Tunable acoustic metalens.**

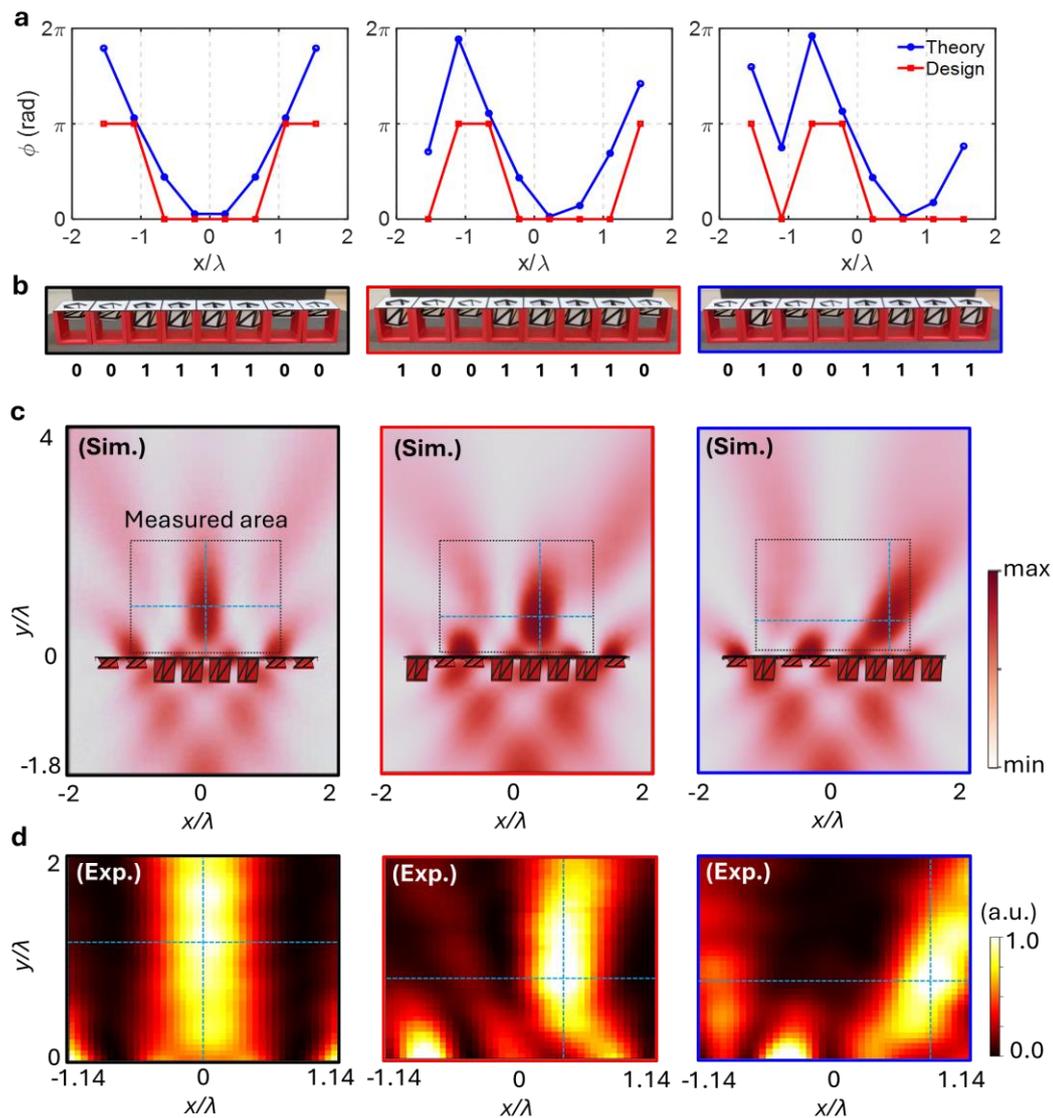

**Figure S5**. Reconfigurable acoustic metalens. a) The theoretical and design phase shifts of the reconfigurable metasurface for different coding sequences, from left to right: 00111100, 10011110, and 01001111. b) The corresponding prototype of the fabricated metasurface. c) Simulation results for the metasurface at the three specified coding sequences. d) Measured results showing the different focal points corresponding to the various coding sequences.

The proposed origami metasurface can be customized to achieve various focusing behaviors with a steerable focal point. To effectively focus sound energy, we initially designed the reflection phase distribution of each origami unit using Generalized Snell's Law (GSL). The theoretical reflection phases of the 8-unit metasurfaces, derived from the GSL, are represented by the blue lines in

Figure S5a. Since the theoretical phase shift covers the range of $[0, 2\pi]$, while the designed phase shift of the origami unit is either $0$ or $\pi$, we adapt the calculated phase shift for each origami unit as follows:

- If the calculated phase shift is larger than $\pi$, the designed phase shift is set to $\pi$, resulting in the origami unit being in its second equilibrium state.

- If the calculated phase shift is less than $\pi$, the designed phase shift is set to $0$, resulting in the origami unit being in its first equilibrium state.

The red line in Figure 5a exhibits the final designed phase shifts for the three metasurface configurations (00111100, 10011110, and 01001111), while Figure S5b shows the fabricated metasurfaces for these configurations. The simulated and measured results of these metasurface configurations, depicted in Figures S5c and S5d, respectively, reveal the metasurface's ability to focus at three specific positions $(0\ \text{mm}, 150\ \text{mm})$, $(60\ \text{mm}, 110\ \text{mm})$, and $(132\ \text{mm}, 100\ \text{mm})$. Figure S6 compares the simulation and measurement of normalized sound intensity at the focal plane. The strong agreement between simulation and experimental results demonstrates that the metasurface can function as a tunable metalens at 2000 Hz by reprogramming its configuration.

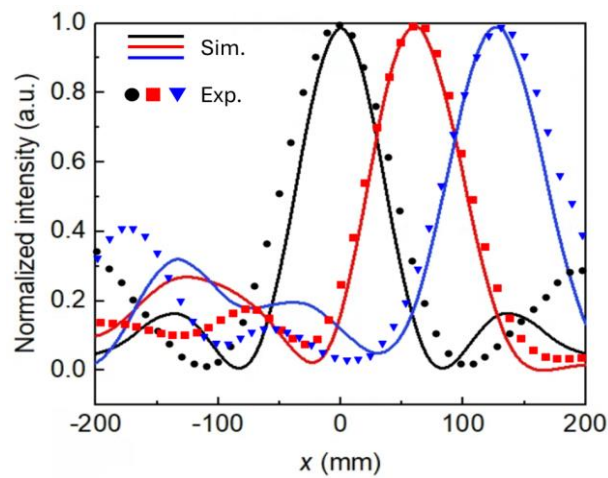

**Figure S6**. Normalized sound intensity for three different configurations of the metalens. Solid lines represent simulation results, while scatter points denote the experiment. The colors black, red, and green correspond to the normalized sound intensity at the focal plane along the x-axis of the metasurface for different coding sequences: 00111100, 10011110, and 01001111, respectively.

**Supplementary Note 5: Steerable acoustic splitter.**

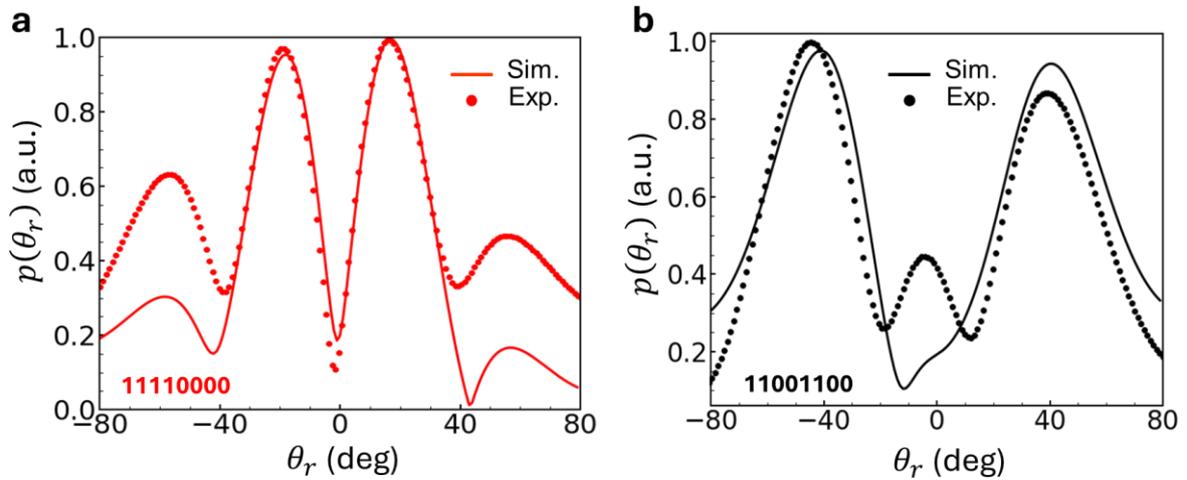

**Figure S7.** Steerable acoustic splitter. Normalized scattering patterns when the coding sequence is a)11110000 and b) 11001100

Figure S7 illustrates the steerable acoustic splitter function of the proposed origami metasurface, demonstrating the substantial agreement between simulation and experimental results for the normalized far-field scattered pressure. With a coding sequence of 11110000, the metasurface generates two symmetric reflection beams at angles of −20° and +20° (Figure S7a), as discussed in the main text. By adjusting the coding sequence to 11001100, the metasurface produces two symmetric reflection beams at −40° and +40°, as presented in Figure S7b. It is worth noting that in this paper, only the 8-origami metasurface is numerical and experimentally investigated due to the physical size of the measurement setup; hence, we demonstrated only two configurations of the metasurface for the acoustic splitting function. The metasurface with more origami units can be customized to have more configuration, producing more reflection angles.

**Supplementary Note 6. Comparison of measured reflection phase between rigid and flexible structures**

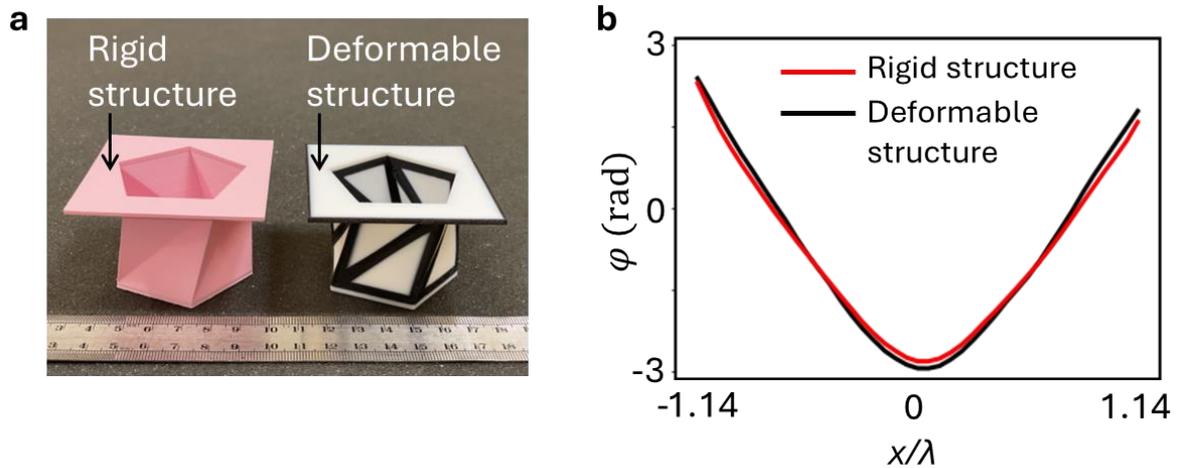

**Figure S8.** a) Photographs of our proposed deformable structure and the rigid counterpart. b) Measured reflection phases between two structures.

We compared the reflection phase of our proposed deformable origami unit with that of its rigid counterpart. The rigid structure was fabricated using polylactic acid (PLA) filament through conventional FDM 3D printing technology, maintaining the same dimensions as the deformable structure, as shown in Figure S8a. The measured reflection phases between the two structures, shown in Figure S8b, depict the excellent agreement between rigid origami and its counterpart. The phase difference is zero ($\Delta\varphi = 0$) at most positions, indicating that the deformable origami unit performs similarly to the rigid one. This similarity in performance results from the comparable mechanical properties of VeroMagentaV and PLA, as well as the relatively small area of flexible material compared to the rigid material (see Figure S2a).